\documentclass[a4paper,twocolumn,notitlepage,twoside,fleqn]{article}

\newcommand{\bra}[1]{\langle {#1} |}
\newcommand{\ket}[1]{| {#1} \rangle}

\newcommand{\vecr}{{\bf r}}
\newcommand{\vecq}{{\bf q}}

\usepackage{proceedings}
\usepackage[dvips]{graphicx}

\HASHIRA{%
\small\textsf{%
RIKEN Review No.39 (???, 2001): {\slshape Focused on}
Physics at Drip Lines
}}

\begin{document}

\title{%
3D real-space calculations of continuum response}

\author{%
Takashi Nakatsukasa,$^{*1}$ and
Kazuhiro Yabana$^{*2}$ \\
{\normalsize $^{*1}$ {\itshape RI Beam Science Laboratory,
 RIKEN, Wako 351-0198, Japan} \\
$^{*2}$ {\itshape Institute of Physics, University of Tsukuba,
 Tennodai 1-1-1, Tsukuba 305-8571, Japan}}\\
}

\abstract{%
We present linear response theories in the continuum capable of
describing continuum spectra and dynamical correlations
of finite systems with no spatial symmetry.
Our formulation is essentially the same as the continuum random-phase
approximation (RPA) but
suitable for
uniform grid representation in the three-dimensional (3D)
Cartesian coordinate.
Effects of the continuum are taken into account by solving equations
iteratively with a retarded Green's function.
The method is applied to photoabsorption spectra in small molecules
(acetylene and ethylene) and inelastic electron scattering from a
deformed nucleus $^{12}$C.
}

\maketitle



\section*{Introduction}
Drip-line nuclei are weakly bound finite fermion systems.
One would naturally expect that the continuum should be taken into
account explicitly in the description of their excited states.
Bound ($L^2$) representation,
which is a popular prescription to construct spatially localized states,
may give the gross properties
of these states embedded in the continuum.
However, the bound representation is incapable of describing
quantities such as escape width and continuum spectra.
The inclusion of the single-particle continuum for particle-hole (p-h) 
excitations has been achieved by using the so-called continuum
random-phase approximation (RPA).$^{1)}$
The continuum RPA combined with the Skyrme Hartree-Fock (HF) theory
has been extensively utilized to study giant resonances (GRs)
in spherical nuclei$^{2,3)}$.
Photoabsorption in rare gases was also studied by using the time-dependent
local density approximation (LDA).$^{4,5)}$
One drawback of the method is that its applicability is limited only to
systems with spherical symmetry.
In this paper, we would like to present a prescription to treat the full
three-dimensional (3D) continuum in the RPA level.

For finite electron systems such as molecules,
theoretical analysis of photoabsorption spectra above the first ionization
threshold requires continuum electronic wave function in a non-spherical
multicenter potential.
Advances in measurements with synchrotron radiation
and high-resolution electron energy loss spectroscopy have enabled us to
obtain oscillator strength distribution of an entire spectral region
originated from valence electrons.$^{6)}$
The photon energy
dependences of molecular photoelectron spectra have also been measured.
The spreading width is less important in molecules than nuclear GR cases;
however, there is a small width associated with molecular vibrations
and dissociations.

Since the Hamiltonian in the Skyrme HF theory for nuclei
and the one in the Kohn-Sham (KS) theory of the
LDA for electron systems are almost diagonal in
coordinate representation, a grid representation in the coordinate space
provides an economical description.
A uniform grid representation in the
3D Cartesian coordinate 
is becoming a standard method for calculation of the ground state.$^{7,8)}$
We would like to use the same 3D basis for calculation of the continuum
response.
The main problem lies in how to incorporate the scattering boundary condition
in the uniform grid representation.

Our method is based on the modified Sternheimer method.$^{9)}$
It recasts the linear response problem into solving the
static Schr\"odinger-type equation with a source term.
The problem is
then to solve the equation with an appropriate outgoing boundary condition.
Our recipe here is to solve iteratively the equation taking into account
the boundary condition employing a Green's function of a spherical potential,
separating the self-consistent potential into long-range spherical and
short-range non-spherical parts.
This is reviewed in the next section.

\section*{Linear response in the continuum of 3D real space}

Exact treatment of the continuum is possible with the use of a
Green's function with an outgoing boundary condition.
For spherical systems, the Green's function can easily
be constructed by making a multipole expansion and discretizing the
radial coordinate.
This is an essential part of the continuum RPA method.
In this section we present a method to construct
a Green's function in the 3D grid representation for a system
without any spatial symmetry.

The linear response theory is formulated most conveniently using a
retarded density-density correlation function.$^{10)}$
\begin{eqnarray}
\Pi(\vecr,\vecr';\omega)&=& \int dt e^{i\omega t-\eta t}
                            \Pi(\vecr,t;\vecr',0) ,\\
\label{Pi_def}
i \Pi(\vecr,t;\vecr',t') &=& \theta(t-t')
  \bra{0} [ \hat{\rho}(\vecr,t),\hat{\rho}(\vecr',t') ] \ket{0} .
\end{eqnarray}
The RPA transition density in response to an external field $V_{\rm ext}$
can be expressed as
\begin{eqnarray}
\delta \rho(\vecr,\omega) &=&
 \int d^3r \Pi_{\mbox{\sc rpa}}(\vecr,\vecr';\omega)
 V_{\rm ext}(\vecr,\omega) \\
\label{delta_n_omega}
 &=& \int d^3 r' \Pi_0(\vecr,\vecr';\omega)
    V_{\rm scf}(\vecr',\omega) ,
\end{eqnarray}
where the $\Pi_0$ is 
the independent-particle density-density correlation function
which is defined by
identifying the state $\ket{0}$ in Eq. (\ref{Pi_def})
with the unperturbed HF (KS) ground state and assuming that
the density operator is evolved in time
with the static HF (KS) Hamiltonian.
$V_{\rm scf}$, the self-consistent field, is the sum of
the external field and the dynamical induced field:
\begin{equation}
\label{Vscf}
V_{\rm scf}(\vecr,\omega) = V_{\rm ext}(\vecr,\omega)
 + \int d^3r' \left.
         \frac{\delta V[\rho(\vecr)]}{\delta \rho(\vecr')}
         \right|_{\rho_0}
         \delta \rho(\vecr';\omega) ,
\end{equation}
where $V[\rho(\vecr)]$ is a single-particle potential of
either the HF or the KS equation.
In order to treat the continuum, we express the $\Pi_0$
in the form$^{1,4)}$
\begin{eqnarray}
\label{Pi_0}
\Pi_0(\vecr,\vecr';\omega) &=&
 \sum_i^{\rm occ}
 \phi_i(\vecr)\left\{
     \left(G^{(+)}(\vecr,\vecr';(\epsilon_i-\omega)^*)\right)^*
   \right. \nonumber \\
   && +\left. G^{(+)}(\vecr,\vecr';\epsilon_i+\omega) \right\}
   \phi_i(\vecr') ,
\end{eqnarray}
assuming that the occupied states have real wave functions $\phi_i$.
$G^{(+)}$ is a single-particle Green's function with an outgoing
asymptotic boundary condition, and is defined by
\begin{eqnarray}
\label{G_sp}
G^{(+)}(\vecr,\vecr';E)
&=& \bra{\vecr}
\left( E + \frac{1}{2m}\nabla^2 - V(\vecr) + i\eta \right)^{-1}\ket{\vecr'}
\\
&=& \sum_k \frac{\phi_k(\vecr)\phi_k^*(\vecr')}{E-\epsilon_k + i\eta},
\end{eqnarray}
where $\phi_k$ and $\epsilon_k$ are the HF (KS) single-particle orbitals and
their eigenenergies, respectively.

In writing the Green's function in terms of single-particle orbitals,
as in Eq. (\ref{G_sp}), we need to calculate both occupied and unoccupied
orbitals including the continuum wave functions with positive energies.
This would be a difficult task.
The trick of the continuum RPA is an explicit construction of the
Green's function in the coordinate space.
If the potential $V(r)$ has spherical symmetry,
$G^{(+)}(\vecr,\vecr';E)$ can be reduced into the radial function
$g_l^{(+)}(r,r';E)$.
Then, we can easily construct $g_l^{(+)}$ as
\begin{equation}
g_l^{(+)}(r,r';E)=2m\frac{u_l(r_<) w_l^{(+)}(r_>)}{W[u_l,w_l^{(+)}]} ,
\label{g_l}
\end{equation}
where $u_l$ and $w_l^{(+)}$ are independent solutions of the radial
HF (KS) equation.
The asymptotic boundary condition of $g_l^{(+)}$
is given by an asymptotic behavior of
the irregular solution $w_l^{(+)}(r)$.
Unfortunately, this explicit construction of the Green's function
is possible only for a one-dimensional case.
It is difficult to impose the boundary condition for a case of
a deformed potential.
Furthermore,
for 3D cases, the number of spatial grid points is
too large to construct the Green's function
($N_{\rm grid}\times N_{\rm grid}$ matrix).

Instead of explicit calculation of the Green's function,
we shall utilize a modified Sternheimer method.$^{9)}$
The transition density can be written in the form
\begin{equation}
\delta\rho(\omega)=
 \sum_i^{\rm occ} \phi_i
\left\{ \left(\psi_i((\epsilon_i-\omega)^*)\right)^*
+ \psi_i(\epsilon_i+\omega) \right\} ,
\end{equation}
where functions $\psi_i$ are defined by 
\begin{equation}
\label{psi_i}
\psi_i(\vecr;E,V_{\rm scf}) \equiv
\int d^3r' G^{(+)}(\vecr,\vecr';E) V_{\rm scf}(\vecr') \phi_i(\vecr') .
\end{equation}
Using properties of the Green's function, we have an equation
\begin{equation}
\label{Sternheimer}
\left(E + \frac{1}{2m}\nabla^2 - V\right) \psi_i(\vecr;E,V_{\rm scf})
    = V_{\rm scf}(\vecr) \phi_i(\vecr) .
\end{equation}
The integral in Eq.~(\ref{psi_i}) is thus converted into a differential
equation (\ref{Sternheimer}).
This procedure is known as the
modified Sternheimer method.$^{9)}$
To determine a solution of Eq. (\ref{Sternheimer}) uniquely,
we have to specify the outgoing boundary condition.
This is our final task.

We assume that the potential $V$ can be divided into two parts,
a long-range spherical part $V_0(r)$ and
a short-range non-spherical part $v=V-V_0(r)$.
For instance, $V_0(r)$ is taken as a Coulomb potential induced
by a spherical charge distribution.
A necessary condition is that $v$ must vanish outside the box
(model space).
Let us denote a Green's function for the Hamiltonian
$H_0=-\nabla^2/2m +V_0$ as $G_0^{(+)}(E)$, and
first calculate an action of $G_0^{(+)}$ on a vector $\Phi$,
$\Psi(E)= G_0(E) \Phi$, by solving the differential
equation
\begin{equation}
\label{dif_eq_G0}
\left\{ E - \left(-\frac{1}{2m}\nabla^2 +V_0(\vecr)\right) \right\}
\Psi(\vecr;E)  = \Phi(\vecr).
\end{equation}
The solution at the boundary of the box ($r\geq R$) is prepared using
a multipole expansion method.
\begin{eqnarray}
\label{Psi}
\left.\Psi(\vecr;E)\right|_{r\geq R} =
 \sum_{lm}^{l_{\rm max}} \frac{w_l^{(+)}(r;E)}{r}
  Y_{lm}(\hat r) \Phi_{lm}(E) ,\\
\label{Phi_lm}
\Phi_{lm}(E) \equiv 2m \int_{r<R} d^3r'
 \frac{u_l(r';E)}{r'} Y_{lm}(\hat r') f(\vecr'),
\end{eqnarray}
where $u_l(r;E)$ and $w_l^{(+)}(r;E)$ are solutions of the radial
differential equation being normalized as the Wronskian
$W[u_l,w_l^{(+)}]$ is unity.
Then, we use the following identity for the Green's function
\begin{equation} G^{(+)}(E)=G_0^{(+)}(E) + G_0^{(+)}(E) v G^{(+)}(E) ,
\end{equation}
to calculate actions of $G^{(+)}(E)$.
Namely, instead of solving Eq. (\ref{Sternheimer}), we solve the equation
\begin{equation}
\label{psi_i_eq}
(1 -G_0^{(+)}(E) v) \psi_i(E) = G_0^{(+)}(E) V_{\rm scf} \phi_i ,
\end{equation}
by using an iterative method.
In this way, we fix an outgoing asymptotic boundary condition for the
Green's function $G^{(+)}(E)$.

Now let us describe a numerical algorithm for
calculation of transition density $\delta\rho$.
There are three-nested iterative loops
($ \mbox{I} \supset \mbox{II} \supset \mbox{III}$)
to solve the multilinear equations:\\
\hspace*{5pt} I) Solve Eq. (\ref{delta_n_omega});
    $(1-\Pi_0 \frac{\delta V}{\delta\rho} ) \delta\rho = \Pi_0 V_{\rm ext}$.\\
\hspace*{5pt}  II) Calculate actions of $G^{(+)}$ by solving Eq. (\ref{psi_i_eq}).\\
\hspace*{5pt}  III-1) Evaluate boundary values of actions of $G_0^{(+)}$
 using
\hspace*{33pt} Eqs. (\ref{Psi}) and (\ref{Phi_lm}).\\
\hspace*{5pt}  III-2) Calculate actions of $G_0^{(+)}$ by solving Eq. (\ref{dif_eq_G0}).\\
We use the generalized conjugate gradient method for non-hermitian problems,
I) and II).
For hermitian problems such as III-2), the conjugate gradient method is
the most efficient algorithm.
In the procedure I), if we neglect the dynamically induced field
$\delta V/\delta\rho$, the calculated response is called
``independent-particle approximation'' (IPA) in the following sections.

\section*{Continuum RPA vs discrete RPA}

In this section, we compare results of the continuum RPA calculation
with those of the RPA discretized in a spherical box.
The modified Skyrme force, so-called ``BKN interaction'',$^{11)}$
is adopted.
\begin{eqnarray}
h_{\mbox{\sc hf}} &=&-\frac{1}{2m}\nabla^2 + \frac{3}{4}t_0\rho(\vecr)
  +\frac{3}{16}t_3\rho(\vecr)^2\nonumber\\
 &&+ V_0 a \int d\vecr' \frac{\exp(-|\vecr-\vecr'|/a)}{|\vecr-\vecr'|}\rho(\vecr')\nonumber\\
 &&+ \frac{e^2}{4} \int d\vecr' \frac{1}{|\vecr-\vecr'|}\rho(\vecr') .
\end{eqnarray}
where spin-isospin degeneracy (each nucleon with a charge $e/2$) is
assumed.
Thus, $\rho(\vecr)=4\sum_i |\phi_i(\vecr) |^2$.
The same parameters as those in Ref. 11) are used in the
calculation.
The self-consistent HF potential is calculated in the spherical box
of radius $R=8$ fm for the $^{16}$O nucleus.
The single-particle energies for $0s$ and $0p$ orbitals are
$-28.4$ and $-16.7$ MeV, respectively.

The strength function is defined by
\begin{eqnarray}
\label{SE_0}
S(E)&=&\sum_n |\bra{n} V_{\rm ext} \ket{0}|^2 \delta(E-E_n) ,\\
\label{SE_1}
    &=& -\frac{1}{\pi} \mbox{Im}\int d\vecr
                         V_{\rm ext}^\dagger(\vecr) \delta\rho(\vecr;E) ,\\
\label{SE_2}
    &=& -\frac{1}{\pi} \mbox{Im Tr}\left( 
         V_{\rm ext}^\dagger \Pi_{\mbox{\sc rpa}}(E) V_{\rm ext} \right).
\end{eqnarray}
Since the obtained HF potential of $^{16}$O is spherical,
we can calculate the
continuum response in the radial coordinate following the conventional
continuum RPA procedure.$^{1)}$
The single-particle Green's function is calculated as Eq. (\ref{g_l})
and the RPA response function can be explicitly constructed by
\begin{equation}
\label{Pi_RPA}
\Pi_{\mbox{\sc rpa}}(E)=
   \left( 1+\Pi_0(E)\frac{\delta V}{\delta\rho} \right)^{-1}
    \Pi_0(E) .
\end{equation}
The strength function is calculated with Eq. (\ref{SE_2}).
The adopted box is a sphere of radius $R=8$ fm
discretized in meshes of $0.1$ fm.
Thus, the response function $\Pi(r,r';E)$ is
a matrix of $80\times 80$.
This size of matrices can be easily handled and the matrix inversion
in Eq. (\ref{Pi_RPA}) can be done.
However, for 3D calculations in the following sections,
since the size of the matrix becomes of order 10 000,
we shall use the Sternheimer method with the outgoing boundary condition
described in the previous section.
The strength function is calculated with Eq. (\ref{SE_1}).

In order to see the effects of the continuum,
we calculate the ``discrete'' RPA response as well.
This can be done by simply using a standing-wave solution
$w_l^{(0)}(r)$ ($w_l^{(0)}(R)=0$) in Eq. (\ref{g_l})
instead of the outgoing solution $w_l^{(+)}(r)$.
Alternatively, we may solve
the modified Sternheimer equation (\ref{Sternheimer})
with a boundary condition $\psi_i=0$ for $r>R$.

\begin{figure}[t]
  \begin{center}
     \includegraphics[height=16.8pc]{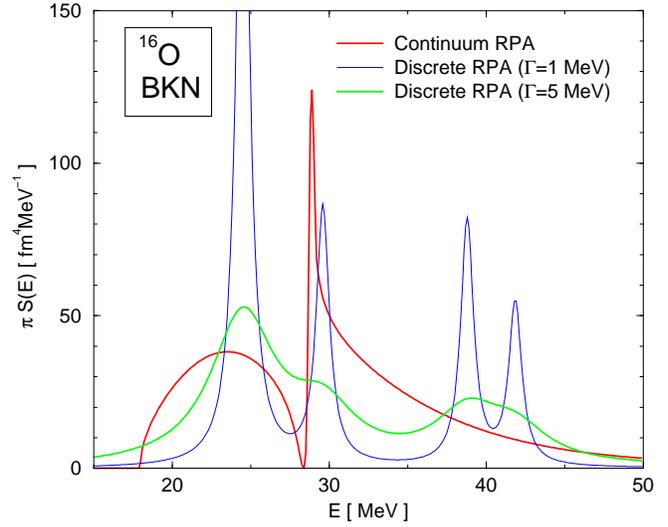}
  \end{center}
  \caption{
Calculated monopole strength distributions for $^{16}$O.
The red line indicates results of the continuum RPA
while the blue (green) line
is the ``discrete'' RPA with a smoothing parameter
$\Gamma$=1 (5) MeV.
}
\label{GMR_O16}
\end{figure}

Calculated strength functions for isoscalar giant monopole resonance
(ISGMR) are shown in Fig.~\ref{GMR_O16}.
The external field is $V_{\rm ext}=r^2$ in this calculation.
For discrete RPA calculation,
since all excited states are discrete, we use complex energies
$E=E-i\Gamma$
to produce a continuous profile of the
strength functions.
As we see in the figure, the continuum response is significantly
different from the smoothed discrete response.
In the discrete calculation, there are peaks characterized by
the size of the box.
In Fig.~\ref{GMR_O16}, concentrated strengths appearing around 40 MeV
are spurious peaks produced by the finite box.
The positions of these spurious peaks will change
as the size of the box is modified.
In the continuum calculation,
a dip emerges at the threshold energy of the $0s$ orbital (28.4 eV)
but there is only a smooth tail beyond 30 MeV.

\section*{Inelastic electron scattering from $^{12}$C}

In this section, we discuss a 3D calculation of the continuum response.
As we have mentioned above, although the continuum RPA has an ability to
treat the single-particle continuum exactly,
it has never been applied to deformed nuclei because of the difficulty
of constructing the Green's function.
Therefore, the continuum effect on deformed nuclei has rarely been
investigated so far.
We use the same BKN interaction as in the previous section and
try the first calculation for a light nucleus $^{12}$C.
The model space is a 3D spherical box of radius $R=8$ fm with square meshes
of $\Delta x=\Delta y=\Delta z=1$ fm.
The obtained HF ground state has an oblate shape in which
the axis ratio is approximately three to two.

\begin{figure}[t]
  \begin{center}
     \includegraphics[width=0.4\textwidth]{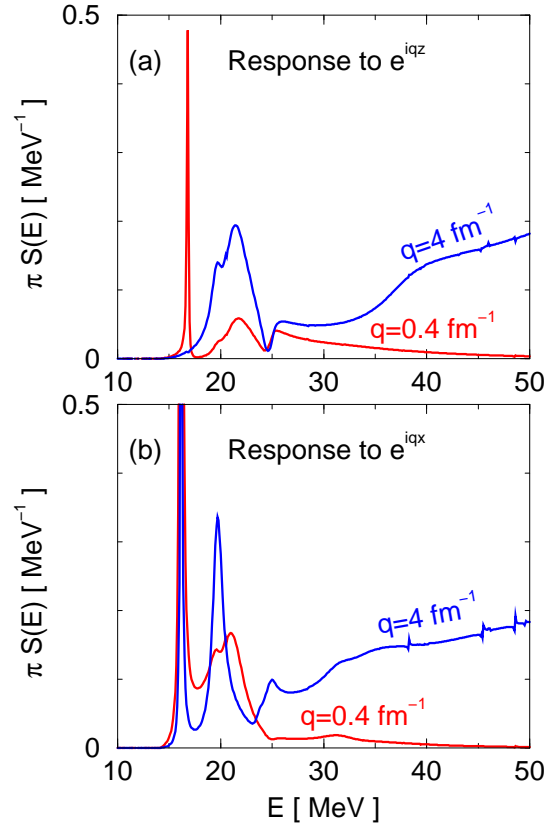}
  \end{center}
  \caption{
Response of $^{12}$C to an electromagnetic probe.
The red and blue lines correspond to momentum transfer q=0.4 and 4
fm$^{-1}$, respectively.
(a) The momentum transfer $\vecq$ is parallel to the symmetry axis of
the nucleus.
(b) The $\vecq$ is perpendicular to the symmetry axis.
}
\label{C12}
\end{figure}

Utilizing the Born approximation,
the cross section of the inelastic electron scattering
for momentum transfer $q$ and energy loss $E$ is approximated by
a product of the strength function and the Mott cross section.$^{10)}$
\begin{equation}
\frac{d\sigma}{d\Omega}=\left.\frac{d\sigma}{d\Omega}
                              \right|_{\rm Mott} S(E,\vecq) ,
\end{equation}
where $S(E,\vecq)$ is defined
by Eqs. (\ref{SE_0}-\ref{SE_2}) with
$V_{\rm ext}=e^{i\vecq\cdot\vecr}(1+\tau_3)/2$.\footnote{
Because of the assumption of the BKN interaction,
the isoscalar response is shown in Fig.~\ref{C12},
with $V_{\rm ext}=e^{i\vecq\cdot\vecr}$.
}
Since the ground state of $^{12}$C is deformed,
the response to $V_{\rm ext}$ depends on the direction of
momentum transfer $\vecq$ relative to the orientation of the nucleus.
We show results for two cases that the $\vecq$ is parallel and perpendicular
to the symmetry axis ($z$-axis) of the nucleus.
Of course, in experiments, the response averaged over different angles
is observed.

In Fig.~\ref{C12}, the strengths as functions of energy are displayed
for low momentum transfer ($q=0.4\mbox{ fm}^{-1}$) and for high momentum
transfer ($q=4\mbox{ fm}^{-1}$).
It seems that peaks related to giant resonances can be seen up to
25 MeV.
For the case of high momentum transfer, the response strongly depends on
the direction of $\vecq$ relative to the intrinsic orientation.

\section*{Photoabsorption of molecules}

First we discuss valence-shell photoabsorption of acetylene and ethylene
molecules.
The KS Hamiltonian for valence electrons is
\begin{equation}
h_{\mbox{\sc ks}} = -\frac{1}{2m}\nabla^2
+ V_{\rm ion}
+ e^2 \int d^3 r' \frac{\rho(\vecr')}{|\vecr - \vecr'|}
+\mu_{\rm xc}[\rho(\vecr)] ,
\end{equation}
where $V_{\rm ion}$ is an electron-ion potential for which we employ
a norm-conserving pseudopotential$^{12)}$ with a separable
approximation.$^{13)}$
The $\mu_{\rm xc}$ is an exchange-correlation potential.

It is well known that the energy of the highest occupied molecular orbital
(HOMO) does
not coincide with the first ionization potential in the simplest
local-density approximation. This fact causes a serious  problem
for the continuum response calculation that the ionization threshold
cannot be adequately described by the static KS Hamiltonian.
Furthermore, the excited states around the ionization threshold appear
in too low excitation energies. To remedy this defect, a gradient
correction for the exchange-correlation potential has been proposed.
We utilize the one constructed by van Leeuwen and Baerends$^{14)}$
which we denote as $\mu^{\rm (LB)}$. It is so constructed that the potential
has a correct $-e^2/r$ tail asymptotically. The energy of the highest
occupied orbital also approximately coincides with the ionization
potential. For small molecules, TDLDA calculations with this gradient
correction have shown to give an accurate description of discrete excitations
in small molecules.$^{15)}$

In the following sections,
we employ a sum of the exchange-correlation potentials
of $\mu^{\rm (PZ)}$ of Ref. 16) for the local-density part and
$\mu^{\rm (LB)}$ for
the gradient correction.
We should remark here that an accurate calculation of the gradient correction
$\mu^{(\rm LB)}(\vecr)$ becomes difficult at far outside the molecule,
because $\mu^{(\rm LB)}$ depends on $|\nabla n(\vecr)|/n(\vecr)^{4/3}$
which approaches a finite value but both the numerator and the
denominator approach zero at $r\rightarrow \infty$.
Thus, we use an explicit asymptotic form,
$\mu^{(\rm LB)}(\vecr)=-e^2/r$ for $r > R_{\rm c}$.
In the following,
$R_{\rm c}$ is chosen as 5.5 \AA.
The spherical box is taken as radius $R=6$ \AA\ with meshes of
$\Delta x=\Delta y=\Delta z=0.3$ \AA.

The acetylene molecule, C$_2$H$_2$,
has a symmetry configuration of $D_{\infty h}$.
This high symmetry has enabled calculation of the Green's function
using a single-center expansion.$^{17)}$
Even so, only two kinds of molecular orbitals, $3\sigma_g$ and $1\pi_u$,
which are primarily derived from
atomic $p$ states, have been taken into account in Ref. 17),
because it was difficult to describe the $s$-derived states in the
single-center formulation.
Since our framework is free from this problem,
we consider all valence orbitals, including
the $2\sigma_g$ and $2\sigma_u$ in addition to the above $p$-derived orbitals,
to calculate the photoresponses.

\begin{figure}[t]
  \begin{center}
     \includegraphics[height=16.8pc]{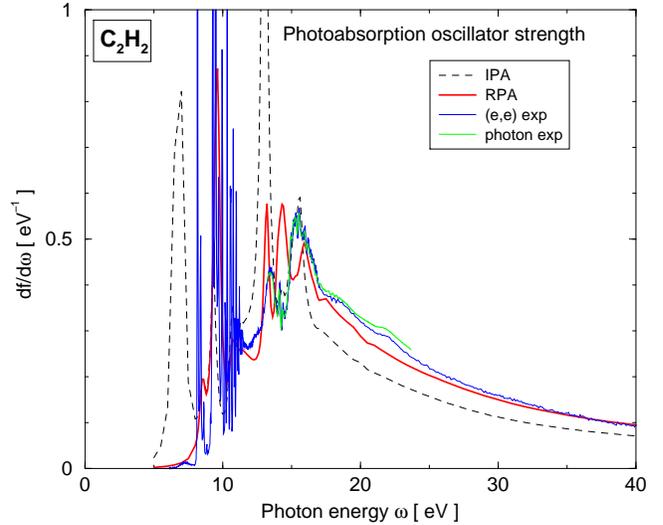}
  \end{center}
  \caption{
Calculated and experimental photoabsorption oscillator strengths
of acetylene.
The red line is the calculation compared with
synchrotron radiation experiment (green line)$^{22)}$ and
high-resolution dipole (e,e') experiment (blue line).$^{23)}$
The dashed line is the IPA calculation without dynamical screening.
}
\label{C2H2_GF}
\end{figure}

All atoms are located on the $z$-axis at $\pm 0.601$ \AA\ for carbon
and $\pm 1.663$ \AA\ for hydrogen.
There are ten valence electrons and
the calculated energies of occupied orbitals are
listed in Table~\ref{IP}.
We obtain the photoabsorption oscillator strengths
as a function of photon energy, as shown in Fig. \ref{C2H2_GF}.
The calculation indicates a sharp bound resonance at $\omega=9.6$ eV
and a broad structure around 15 eV which seems to be a superposition of
three resonances.
The resonance at 9.6 eV strongly responds to a dipole field parallel
to the molecular axis.
The large oscillator strengths in the IPA at $\omega=5\sim 8$ eV and
at $\omega=12.5\sim 13.5$ eV
are shifted to higher energies by the dielectric correlation.
The agreement with the experimental data is significantly improved
by the inclusion of the dynamical induced field (screening).
The static dipole polarizability is also affected significantly:
In the IPA calculation,
the polarizabilities parallel ($\alpha_\parallel$) and perpendicular
($\alpha_\perp$) to the molecular axis are $\alpha_\parallel=10.7$ \AA$^3$
and $\alpha_\perp=3.87$ \AA$^3$.
The dynamical screening reduces these values to 4.79 \AA$^3$ and 2.77 \AA$^3$,
respectively, which well agree with the experimental values,
$\alpha_\parallel=4.73$ and $\alpha_\perp=2.87$ \AA$^3$.$^{18)}$

Another example we show here is ethylene, C$_2$H$_4$.
Ethylene is the simplest organic $\pi$-system,
possessing the $D_{2h}$ symmetry.
Its two carbon atoms lie on the $x$-axis at $x=\pm 0.6695$ \AA\
and its four hydrogen atoms lie in the $x-y$ plane at
$(x,y)=(1.2342,0.9288)$, $(1.2342,-0.9288)$, $(-1.2342, 0.9288)$,
$(-1.2342,-0.9288)$ in units of \AA.
There are 12 valence electrons, and
calculated eigenenergies of occupied orbitals are listed in Table~\ref{IP}.

\begin{figure}[t]
  \begin{center}
     \includegraphics[height=16.8pc]{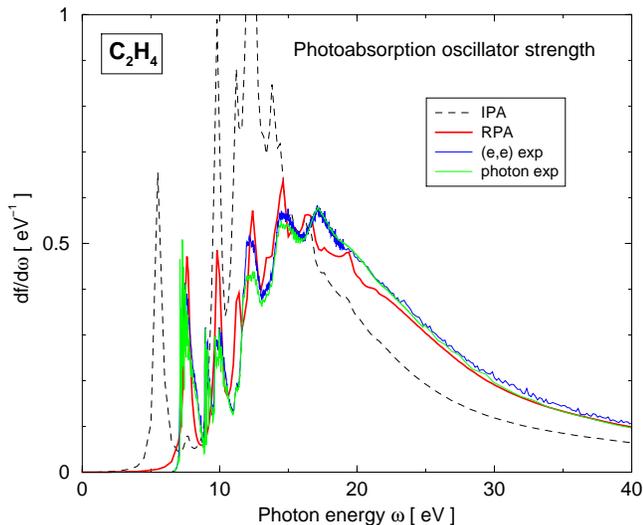}
  \end{center}
  \caption{
The same as Fig.~\ref{C2H2_GF} but for ethylene.
The experimental data are taken from Refs. 19,21).
}
\label{C2H4_GF}
\end{figure}

The calculated photoabsorption oscillator strengths
are shown in Fig. \ref{C2H4_GF}.
The agreement with an experiment$^{19,21)}$ is excellent.
Almost all the main features of photoabsorption spectra
are reproduced in the calculation.
The observed bound excited states show
different photoresponses according to
the direction of the dipole field.
The lowest peak at $\omega=7.6$ eV is mainly a response to a
dipole field parallel to the molecular (C$-$C) axis.
This is associated with the excitations of the HOMO $1b_{3u}$ electrons.
On the other hand, states at 9.8 eV respond almost equally
to a dipole field of $x$, $y$, and $z$ directions,
to which both $1b_{3g}$ and $1b_{3u}$ occupied orbitals contribute.
A small peak at $\omega=11.4$ eV is calculated as a resonance of
$3a_g$ orbital,
which may correspond to a small shoulder in the experiment.
Beyond 11.7 eV, the HOMO electrons are in the continuum.
The first prominent peak at 12.4 eV is a resonance with respect to
a dipole field of $y$ direction which is in the molecular plane
and perpendicular to the C$-$C bond.
The excitations of $1b_{2u}$ and $1b_{3g}$ electrons are the main
components of this resonance.
In the region of $13.2<\omega<20$ eV,
$1b_{3g}$ electrons can be excited into the continuum and
produce the smooth background of oscillator strengths
($0.1\sim 0.2$ eV$^{-1}$).
The peak structure at 14.6 eV is produced by
the excitation of $1b_{2u}$ electrons.
The peak at 16.4 eV is constructed by excitations from
the $2b_{1u}$ occupied orbitals, while the experiment indicates the
resonance at 17.1 eV.

In Figs.~\ref{C2H2_GF} and \ref{C2H4_GF}, we show about 75 \% of
the Thomas-Reiche-Kuhn (TRK) sum
rule for valence electrons.
In other words, one-fourth of the TRK sum rule value lies in an energy
range of over 40 eV.

\newcommand\s{\ \ \ \ }
\begin{table}[htb]
\caption{
Calculated eigenvalues of occupied valence orbitals and
experimental vertical ionization potential (IP) in units of eV.
The experimental data are taken
from Ref. 20) for acetylene and from
Ref. 21) for ethylene.
}
\begin{center}
\begin{tabular}{cccccc}
\multicolumn{3}{c}{Acetylene}
                        & \multicolumn{3}{c}{Ethylene} \\
\hline
occ & cal & exp & occ & cal & exp \\
state &    & IP  & state &    & IP  \\
\hline
$(2\sigma_g)^2$ & $-22.4$ & $ 23.5$ &
 $(2a_g)^2$ & $-22.8$ & $ 23.7$ \\
$(2\sigma_u)^2$ & $-18.4$ & $ 18.4$ &
 $(2b_{1u})^2$ & $-18.6$ & $ 19.4$ \\
$(3\sigma_g)^2$ & $-16.7$ & $ 16.4$ &
 $(1b_{2u})^2$ & $-16.3$ & $ 16.3$ \\
$(1\pi_u)^4$    & $-12.1$ & $ 11.4$ &
 $(3a_g)^2$ & $-14.7$ & $ 14.9$ \\
            &         &         &
 $(1b_{3g})^2$ & $-13.2$ & $ 13.0$ \\
            &         &         &
 $(1b_{3u})^2$ & $-11.7$ & $ 11.0$ \\
\hline
\end{tabular}
\end{center}
\label{IP}
\end{table}

\section*{Conclusions}

We have developed a method based on the time-dependent density-functional
theory of investigating
responses in the continuum for systems with no spatial symmetry.
The method allows us to treat the escaping width and the dynamical
correlation effects self-consistently.
We have shown the ISGMR in the continuum for $^{16}$O 
and compared the results of the continuum and discrete calculations.
The continuum response cannot be reproduced by simply smoothing
the discrete strengths.
Perhaps, this is because the continuum level density is not properly
taken into account in the discrete calculation.
Then,
we have shown a calculation of electromagnetic response function for
a nucleus $^{12}$C and photoabsorption spectra for molecules, acetylene
and ethylene.
The main difficulty is the heavy computational task, especially for
calculations of response near the threshold energies.
Using a single CPU of a Fujitsu VPP700E supercomputer at RIKEN,
the numerical calculation of photoabsorption spectra of ethylene
in Fig.~\ref{C2H4_GF} takes about 30 minutes at a single photon energy
$\omega$.
Since we need to calculate the response to dipole field of the $x$, $y$,
and $z$ directions, it takes about 1.5 hours to calculate a single energy
point.
However, we have found that the inclusion of the small imaginary part
in energy, $E+i\Gamma$, facilitates
convergence of iterative procedures in the calculation.
$\Gamma$ also plays the role of lowering an energy resolution of
the calculations.
Thus, we can choose a value of $\Gamma$ depending on the energy resolution
required in each problem.

The method is capable of calculating
response functions of many-particle systems,
below, near and above the separation energy (ionization threshold)
in a unified manner.
The results seem very promising
and encourage us to apply the method to responses of drip-line
nuclei in the near future.

\section*{References}
\footnotesize
\begin{enumerate}
\item S.~Shlomo and G.~Bertsch, Nucl. Phys. {\bf A243}, 507 (1975).
\item N.~van~Giai and H.~Sagawa, Nucl. Phys. {\bf A371}, 1 (1981).
\item I.~Hamamoto, H.~Sagawa, and X.~Z.~Zhang, Phys. Rev. C {\bf 53},
 765 (1996); I.~Hamamoto and H.~Sagawa, Phys. Rev. C {\bf 53}, R1492 (1996);
 Phys. Rev. C {\bf 54}, 2369 (1996).
\item A.~Zangwill and P.~Soven, Phys. Rev. A {\bf 21}, 1561 (1980).
\item M.~J.~Stott and E.~Zaremba, Phys. Rev. A {\bf 21}, 12 (1980).
\item Y.~Hatano, Phys. Rep. {\bf 313}, 109 (1999).
\item H.~Flocard, S.~E.~Koonin, and M.~S.~Weiss, Phys. Rev. C {\bf 17},
 1682 (1978);
 K.~T.~R.~Davies, H.~Flocard, S.~Krieger, and M.~S.~Weiss,
 Nucl. Phys. {\bf A342}, 111 (1980).
\item J.~R.~Chelikowsky, N.~Troullier, K.~Wu, and Y.~Saad,
 Phys. Rev. B {\bf 50}, 11355 (1994);
 T.~L.~Beck, Rev. Mod. Phys. {\bf 72}, 1041 (2000)
 and references therein.
\item G.~D.~Mahan, Phys. Rev. A {\bf 22}, 1780 (1980).
\item A.~L.~Fetter and J.~D.~Walecka, {\it Quantum Theory of Many-Body Systems}
 (McGraw-Hill, New York 1971).
\item P.~Bonche, S.~Koonin, and J.~W.~Negele, Phys. Rev. C {\bf13},
 1226 (1976).
\item N.~Troullier and J.~L.~Martins, Phys. Rev. B {\bf 43}, 1993 (1991).
\item L.~Kreinman and D.~Bylander, Phys. Rev. Lett. {\bf 48}, 1425 (1982).
\item R.~van~Leeuwen and E.~J.~Baerends, Phys. Rev. A {\bf 49}, 2421 (1994).
\item M.~E.~Casida, C.~Jamorski, K.~C.~Casida, and D.~R.~Salahub,
 J. Chem. Phys. {\bf 108}, 4439 (1998).
\item J.~Perdew and A.~Zunger, Phys. Rev. B {\bf 23}, 5048 (1981).
\item Z.~H.~Levine and P.~Soven, Phys. Rev. A {\bf 29}, 625 (1984).
\item N.~J.~Bridge and A.~D.~Buchkingham, Proc. R. Soc. A {\bf 295},
 334 (1966).
\item G.~Cooper, T.~N.~Olney, and C.~E.~Brion,
 J. Chem. Phys. {\bf 194}, 175 (1995).
\item M.~Ukai, K.~Kameta, R.~Chiba, K.~Nagano, N.~Kouchi, K.~Shinsaka,
 Y.~Hatano, H.~Umemoto, Y.~Ito, and K.~Tanaka,
 J. Chem. Phys. {\bf 95}, 4142 (1991).
\item D.~M.~P.~Holland, D.~A.~Shaw, M.~A.~Hayes, L.~G.~Shpinkova, E.~E.~Rennie,
 L.~Karlsson, P.~Baltzer, and B.~Wannberg, Chem. Phys. {\bf 219}, 91 (1997).
\item K.~Kameta, M.~Ukai, R.~Chiba, K.~Nagano, N.~Kouchi, Y.~Hatano,
 and K.~Tanaka, J. Chem. Phys. {\bf 95}, 1456 (1991).
\item G.~Cooper, G.~R.~Burton, W.~Fat~Chan, and C.~E.~Brion,
 Chem. Phys. {\bf 196}, 293 (1995).
\end{enumerate}

\end{document}